\documentstyle[11pt,paspconf,epsf]{article}
\input epsf

\begin{document}

\title{The emission-line spectrum of KUG 1031+398 and the Intermediate 
Line Region controversy}

\author{Anabela C. Gon\c{c}alves, Philippe V\'eron and Marie-Paule 
V\'eron-Cetty}
\affil{Centre National de la Recherche Scientifique, 
Observatoire de Haute-Provence, 04870 St. Michel l'Observatoire, France}

\begin{abstract}
We present results based on the analysis of optical spectra of the 
Narrow-Line Seyfert 1 (NLS1) galaxy KUG~1031+398, for which evidence was  
reported of a line-emitting region ``intermediate'' (both in terms of 
velocity and density) between the conventional Broad and Narrow Line Regions 
(BLR and NLR, respectively). From our observations and modeling of the 
spectra, we get a consistent decomposition of the line profiles into four 
components: an extended H\,{\sc ii} region with unresolved lines, two distinct 
Seyfert-type clouds identified with the NLR, and a relatively narrow 
``broad line'' component emitting only Balmer lines but no forbidden lines. 
Therefore, although we find this object to be exceptional in having 
line-emission from the BLR with almost the same width as the narrow lines, 
our interpretation of the data does not support the existence of an 
``intermediate'' line region (ILR).
\end{abstract}


\section{Introduction}
\subsection{The Intermediate Line Region}
It is commonly accepted that line-emission in AGN comes from two well 
separated regions: one, compact ($<$ 1 pc) and lying close to the central 
engine, has a high electron density ($N_{\rm e}$ $>$ 10$^{8}$ cm$^{-3}$) 
and is responsible for the production of broad (FHWM $\sim$ thousands of 
km\,s$^{-1}$) permitted lines---the BLR; the other, more extended and lying 
further away from the central source (10--1\,000~pc), has lower electron 
densities ($N_{\rm e}$ $\sim$ 10$^{5}$ cm$^{-3}$) and emits lines with a lower 
velocity dispersion ($\sim$ hundreds of km\,s$^{-1}$)---the NLR. A ``gap'' of 
line-emission is usually observed between the two regions; most objects show 
optical spectra which can be fitted by line profiles corresponding to 
clouds belonging to one or the other line-emitting regions. This line-emission 
gap can be explained by the presence of dust mixed up with the gas (Netzer \& 
Laor 1993). Nevertheless, the existence of an intermediate region, both in 
terms of velocity and density ($N_{\rm e}$ $\sim$ 10$^{6.5}$ cm$^{-3}$) is 
expected; at this density, the [O\,{\sc iii}] lines are partially collisionally 
de-excited such that $\lambda$5007$/$H$\beta$ $\sim$ 1 [this ILR should not 
be confused with the ILR found in QSOs by Brotherton et al. (1994), which is 
much smaller and denser, with velocity dispersion of the order of 2\,000 
km\,s$^{-1}$ and density $\sim$ 10$^{10}$ cm$^{-3}$]. To the best of our 
knowledge, no uncontroversial report of the existence of an ILR has ever been 
made. Crenshaw \& Peterson (1986) and Van Groningen \& de Bruyn (1989) have 
found broad wings in the [O\,{\sc iii}] lines of a number of Seyfert 1 galaxies, 
implying the presence of an ILR in these objects; however, they all show 
strong Fe~{\sc ii} emission, and the observed broad [O\,{\sc iii}] components could be due 
to the inaccurate removal of the Fe\,{\sc ii} blends (Boroson \& Green 1992). The 
claim by Mason et al. (1996) that KUG 1031+398 showed evidence for an ILR 
induced us to conduct new spectroscopic observations and modeling of its 
emission-line features.

\subsection{KUG 1031+398}
KUG 1031+398 has an unusual X-ray spectrum with a very strong soft X-ray 
excess (Pounds et al. 1995; Puchnarewicz et al. 1995); the broad component 
of the Balmer lines is relatively narrow (FWHM $\sim$ 1\,500 km\,s$^{-1}$) 
and, consequently, this object has been classified as a NLS1 by Puchnarewicz 
et al. (1995). However, our spectra do not show strong Fe\,{\sc ii} lines, another 
characteristic of many NLS1.

\section{Observations}
Observations were made with the spectrograph CARELEC (Lema\^{\i}tre et al. 
1989), attached to the OHP 1.93 m telescope. The resolution, as measured on 
the night-sky lines, was $\sim$ 3.4 \AA\ for the blue and $\sim$ 3.5 \AA\ for 
the red spectral regions; the spectral ranges were 
$\lambda\lambda$\,4780--5780 \AA\ and $\lambda\lambda$\,6175--7075 \AA, 
respectively. The slit width was 2\farcs1. Our spectra were flux calibrated 
using the standard stars EG 247 (Oke 1974) and Feige 66 (Massey et al. 1988), 
also used to correct the red spectrum for the atmospheric B band at 
$\lambda$6867 \AA. 

\section{Data Analysis}
Assuming that the emission-line profiles observed in KUG 1031+398 are the 
result of the contributions from several clouds, we tried to model the observed 
spectra with the smallest possible number of line sets, each set including 
three Gaussians (modeling H$\alpha$ and the [N\,{\sc ii}] lines, or H$\beta$ and 
the [O\,{\sc iii}] lines) having the same velocity shift and width, with the 
additional constraint that the intensity ratio of the two [N\,{\sc ii}] 
(respectively [O\,{\sc iii}]) lines was taken to be equal to the theoretical value 
of 3 (respectively 2.96) (Osterbrock 1974). In a physically meaningful and 
self-consistent model, the components found when fitting the blue and red 
spectra should have velocity shifts and widths compatible within the 
measurement errors.
                                           
The spectra were de-redshif\-ted assuming $z$ = 0.0434 and analyzed in terms 
of Gaussian components as described above. We discovered first that the core 
of the lines could not be fitted by a single set of narrow Gaussian profiles. 
To get a satisfactory fit, two sets of Gaussian components are needed: the 
first, unresolved (and subsequently taken as the origin of the velocity 
scales) has $\lambda$6583$/$H$\alpha$ = 0.55, 
$\lambda$5007$/$H$\beta$ = 1.27, and corresponds to a H\,{\sc ii} region; the second 
is resolved (FWHM $\sim$ 350 km\,s$^{-1}$, corrected for the instrumental 
broadening), blueshif\-ted by $\sim$ 95 km\,s$^{-1}$ with respect to the 
narrow components and has line intensity ratios typical of a Seyfert 2 
($\lambda$6583$/$H$\alpha$ = 0.84, $\lambda$5007$/$H$\beta$ $>$ 10).

\begin{figure}
\plotfiddle{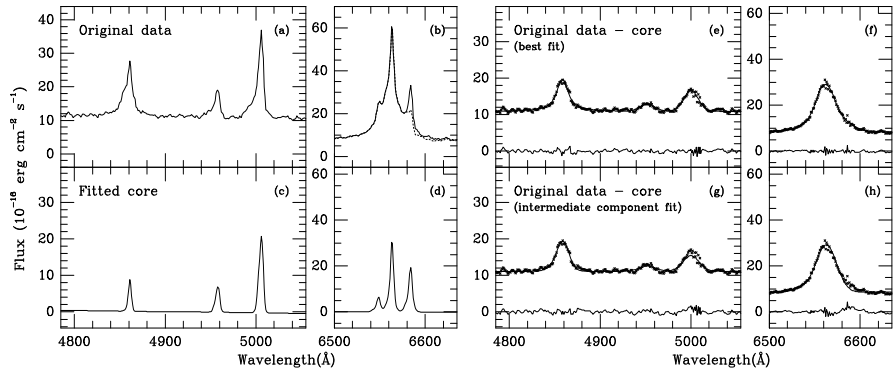}{5cm}{0}{95}{100}{-150}{-100}
\caption{Blue (a) and red (b) spectra of KUG 1031+398 in the rest frame;
in (b) we also give the spectrum before correcting for the atmospheric
absorption (dotted line). The narrow core components (c and d) were fitted
with Gaussians and subtracted from the original data, the result being shown
in (e) to (h). In (e) and (f), we show our best fit (solid line) together
with the data points (crosses); the lower solid lines represent the residuals.
In (g) and (h) we give the fit and residuals obtained when an ``intermediate''
component is imposed, as described in the text.}
\label{Obs_and_Fits}
\end{figure}

At this stage, we removed from the blue and red spectra the best fitting core 
(the H\,{\sc ii} region and the Seyfert 2 nebulosity, Fig. 1c and d), obtaining two 
spectra we shall call ``original data $-$ core''. The blue one was then fitted 
with a broad H$\beta$ Gaussian component and two sets of three components 
modeling the narrow H$\beta$ and [O\,{\sc iii}] lines. The 
result is very suggestive: one set has a strong H$\beta$ line and very weak 
negative [O\,{\sc iii}] components, while 
the other set displays a strong [O\,{\sc iii}] 
contribution and a weak negative H$\beta$ component, showing that we have 
in fact a H$\beta$ component with no 
associated [O\,{\sc iii}] emission and [O\,{\sc iii}] 
lines with a very weak (undetected) associated H$\beta$; in other words, the 
region producing the H$\beta$ line does not emit forbidden lines, while the 
[O\,{\sc iii}] emitting region has a high $\lambda$5007$/$H$\beta$ ratio, which are 
the characteristics of the ``broad'' and ``narrow'' line regions in Seyfert~1 
galaxies, respectively. 

With these results in mind, we optimized this last fit by using a Lorentzian 
profile for the H$\beta$ line, with no associated [O\,{\sc iii}] emission, and a set 
of three Gaussians for the remaining contribution coming from the ``narrow'' 
components; to avoid an unphysical negative intensity for the H$\beta$ line, 
we forced $\lambda$5007$/$H$\beta$ to be equal to 10, which is the ratio 
usually found for the narrow component in Seyfert galaxies. The best fit is 
presented in Fig. 1e. The H$\beta$ Lorentzian component is blueshif\-ted by 
160 km\,s$^{-1}$ with a wi\-dth of 920 km\,s$^{-1}$; the [O\,{\sc iii}] lines are 
blueshif\-ted by $\sim$ 395 km\,s$^{-1}$ and their width is $\sim$ 1\,120 
km\,s$^{-1}$. 

We have also analyzed the ``original data $-$ core'' red spectrum (Fig. 1f)
with one Lorentzian H$\alpha$ component and a set of three Gaussians (for the 
H$\alpha$ and [N\,{\sc ii}] lines) with the constraint that 
$\lambda$6583$/$H$\alpha$ = 0.9, for which we have found a FWHM of $\sim$ 
770 km\,s$^{-1}$ and a blueshift of 375 km\,s$^{-1}$. The H$\alpha$ Lorentzian 
component, blueshif\-ted by 55 km\,s$^{-1}$, has a width of 1\,030 
km\,s$^{-1}$, in reasonable agreement with the width of the corresponding 
H$\beta$ component. The Lorentzian Balmer components, without any measurable 
associated forbidden line, would qualify KUG 1031+398 as a NLS1 with, in fact, 
very narrow lines. The other system of lines, with a very high 
$\lambda$5007$/$H$\beta$ ratio, $\lambda$6583$/$H$\alpha$ $\sim$ 0.9 and 
FWHM $\sim$ 945 km\,s$^{-1}$, is analogous to what is usually found in 
Seyfert~2s and corresponds to a NLR cloud.

At last, we fitted the ``original data $-$ core'' blue spectrum with a 
broad H$\beta$ Gaussian component and one set of three Gaussians (modeling 
H$\beta$ and the [O\,{\sc iii}] lines) for which we set the $\lambda$5007$/$H$\beta$ 
ratio to the value found by Mason et al. for the ``intermediate'' component, 
i.e., 1.42. The red spectrum was fitted with two H$\alpha$ components, for 
which we fixed the redshifts to the values obtained in the blue spectrum 
profile fitting analysis. The resulting fits and residuals, shown in Figs. 1g 
and h, seem to be significantly worse than the ones given in Figs. 1e and f, 
showing that the presence of an ``intermediate'' component is not required 
by the data.

\section{Results and Discussion} 
Our new observations and modeling of KUG 1031+398 yield a consistent 
decomposition of the emission-line profile into four components: an extended 
H\,{\sc ii} region with unresolved lines; a first Seyfert-type cloud with relatively 
narrow lines ($\sim$ 350 km\,s$^{-1}$ FWHM), blueshif\-ted by 95 km\,s$^{-1}$, 
belonging to the NLR; a second Seyfert-type cloud with somewhat broader lines, 
blueshif\-ted by $\sim$ 385 km\,s$^{-1}$, also characteristic of the NLR; and 
finally, a Narrow-Line Seyfert~1 cloud with lines well fitted by a Lorentzian 
profile of $\sim$ 975 km\,s$^{-1}$ FWHM, blueshif\-ted by 105 km\,s$^{-1}$ 
(Table 1).

\begin{table}
\begin{center}
\caption{Emission-line profile analysis of KUG 1031+398. I(H$\beta$) and
I(H$\alpha$) are in units of 10$^{-16}$ erg\,s$^{-1}$\,cm$^{-2}$. The FWHMs
are corrected for the instrumental broadening.} 
\vspace{0.5cm} 
\begin{tabular}{p{0.15cm}rrrcrr}
\tableline
 & $\Delta$V\verb+  + & FWHM\,    &  \underline{$\lambda$5007\,}
 & \underline{$\lambda$6583\,}    &  I(H$\beta$) & I(H$\alpha$) \\
 & (km\,s$^{-1}$) & (km\,s$^{-1}$) & H$\beta$\verb+ +  &  ~H$\alpha$ &  &  \\
\tableline
1       &   0\verb+  +       &   $<$80\verb+ +      &  1.27           &
0.55    &   29$\:\,$         &   93\verb+ + \\
2       &   $-$ 95\verb+  +  &   350\verb+ +        & $>$10.0\verb+ + &
0.84    &   $<$9$\:\,$       &   81\verb+ + \\
3       &   $-$ 385\verb+  + &   $\sim$ 945\verb+ + & (10.0)$\;$      &
(0.90)  &   (11)             &   68\verb+ + \\
4       &   $-$ 105\verb+  + &   975\verb+ +        & --\verb+  +     &
--      &   199$\;$          &   677\verb+ + \\
\tableline
\tableline
\end{tabular}
\end{center}
\end{table}

The model used by Mason et al. to fit the emission-line spectrum of 
KUG 1031+398 seems to support the existence of an ILR emitting lines of 
intermediate width (FWHM $\sim$ 1\,000 km\,s$^{-1}$); this component 
dominates the Balmer line profiles, being also a significant contributor 
to the [O\,{\sc iii}] lines, with a flux ratio $\lambda$5007$/$H$\beta$ = 1.4. 

There are two main reasons why our analysis yields different results. First,  
KUG 1031+398 having a redshift of $\sim$ 0.043, the [N\,{\sc ii}]$\lambda$6583 line 
coincides with the atmospheric B band. When correcting for this absorption 
feature, the [N\,{\sc ii}] true intensity is recovered (Fig. 1b) and our red 
spectrum appears different from the published one; different line-ratios 
and widths are therefore not unexpected.

Second, the line-profile analysis of Mason et al. differs from ours in that, 
while we force each Balmer component to be associated with forbidden lines 
having the same velocity and width, Mason et al. allow these parameters to 
have different values for the Balmer and forbidden line components. As a 
result, they found three H$\beta$ components (a narrow, an intermediate and a 
broad one), as well as two [O\,{\sc iii}] components (a narrow and an intermediate 
one); they also detected three H$\alpha$ components (again a narrow, an 
intermediate and a broad one), but only a single [N\,{\sc ii}] component (narrow). 
The measured width of the narrow H$\beta$ component is 150 $\pm$ 20 
km s$^{-1}$ FWHM, while the width of the narrow [O\,{\sc iii}] lines is 265 $\pm$ 10 
km s$^{-1}$; this last value, significantly larger than the narrow H$\beta$ 
line width, suggests that the [O\,{\sc iii}] lines may have a complex profile. 
Moreover, the width of the [N\,{\sc ii}] lines is found to be significantly larger 
(400 $\pm$ 60 km s$^{-1}$) than that of the narrow H$\alpha$ component 
(190 $\pm$ 40 km s$^{-1}$); this could be due to an inaccurate correction of 
the atmospheric B band, as we have seen before. 
                     
Although our spectra have a lower resolution than those obtained by Mason et 
al. (3.4 \AA\ compared to 2 \AA\ FWHM), this does not affect the analysis; 
the narrow core components being identified and subtracted, all the discussion 
is centered on the broader components, well resolved even with our lower 
resolution. Similarly, the larger slit width used in our observations 
(2\farcs1 compared to 1\farcs5 for Mason et al.) does not affect the study 
of these broader components, since only the contribution from the extended 
emitting region (the H\,{\sc ii} region), removed with the core, changes with the 
slit width.

We disagree with Mason et al. on the result of the line profile analysis of 
KUG 1031+398, in the sense that we find no evidence for the presence of an 
ILR. Nevertheless, we find that this object is exceptional in having a NLR 
(defined as a region where $\lambda$5007$/$H$\beta$ $\ge$ 5) with almost 
the same width as the BLR (Balmer lines with no detectable associated 
forbidden lines).

%
Several authors have suggested that the small width of the broad Balmer 
lines and the soft X-ray excess characteristic of the NLS1 galaxies could be 
the effect of a high accretion rate onto an abnormally small mass black hole. 
Mason et al. (1996) have argued that, although the emission line spectrum in 
this object is dominated by the ILR, a weak broad component is present with 
line-widths of the order of 2\,500 km\,s$^{-1}$ FWHM and that, therefore, at 
least in this object, such a model is not required. Our analysis of the 
spectra shows that in the BLR, the Balmer lines are well fitted by a 
Lorentzian profile with $\sim$ 1\,000 km\,s$^{-1}$ FWHM. We have shown 
(Gon\c{c}alves et al. in preparation) that in NLS1s the broad component of 
the Balmer lines is generally better fitted by a Lorentzian than by a Gaussian; 
so, in this respect, KUG 1031+398 is a normal NLS1 and could be explained by 
the same small black hole mass model suggested for the 
other objects of the same class.

\acknowledgments
Anabela C. Gon\c{c}alves acknowledges support from the {\it Funda\c{c}\~ao 
para a Ci\^encia e a Tecnologia}, Portugal, during the course of this work 
(PRAXIS XXI/BD/5117/95 PhD. grant).

\begin{question}{Jack Sulentic}
Can we agree that there are three types of H$\beta$ line profiles: a) BLR + 
NLR components with clear inflection between them; b) Essentially continuous 
emission at all profile widths between BLR + NLR (this ``ILR'' emission is 
reflected in NLR [O\,{\sc iii}] line profiles); c) same as b) except ``ILR'' emission 
not reflected in NLR [O\,{\sc iii}] ?
\end{question}
\begin{answer}{Anabela Gon\c{c}alves}
The classical Seyfert 1 galaxies usually have a H$\beta$ line profile of 
type ``a''. The NLS1s, which have a broad H$\beta$ component only slightly 
broader than the narrow component, fall into class ``c''. Seyfert 1 galaxies 
with a genuine ILR would be in class ``b''; however, no ILR has yet been 
found. Nevertheless, b-type profiles are sometimes observed in objects with 
strong Fe\,{\sc ii} emission which blends with the [O\,{\sc iii}] and H$\beta$ lines, 
mimicking an ``ILR'' component. Moreover, a number of objects exist which have 
a ``composite'' spectrum, i.e., a H\,{\sc ii} nebulosity and a Seyfert- or 
Liner-like cloud unresolved on the spectrograph slit; in these objects the 
[O\,{\sc iii}] lines can be broader than H$\beta$ (V\'eron et al. 1997), but they 
generally do not have a broad H$\beta$ component. 
\end{answer}

\end{document}